\documentclass[sigconf]{acmart}
\AtBeginDocument{%
  }

\copyrightyear{2026}
\acmYear{2026}
\setcopyright{cc}
\setcctype{by}
\acmConference[MM '26]{Proceedings of the 34th ACM International Conference on Multimedia}{November 10--14, 2026}{Rio de Janeiro, Brazil}
\acmBooktitle{Proceedings of the 34th ACM International Conference on Multimedia (MM '26), November 10--14, 2026, Rio de Janeiro, Brazil}
\acmDOI{10.1145/3767308.3835615}
\acmISBN{979-8-4007-2213-4/2026/11}

\usepackage{multirow} 
\usepackage{graphicx} 
\usepackage{algorithm} 
\usepackage{algpseudocode}
\usepackage[table]{xcolor} % 表格单元格上色核心包，table选项必须加
\usepackage{colortbl}      % 增强表格配色兼容性，可选但推荐加
\usepackage{circledsteps} 
\usepackage[normalem]{ulem}
 \usepackage{balance}
 \usepackage{makecell}
\usepackage{booktabs}
\usepackage{multirow}
\usepackage{adjustbox}
\usepackage{caption}
\definecolor{emerald}{RGB}{0, 128, 104}
\begin{document}

%%
%% The "title" command has an optional parameter,
%% allowing the author to define a "short title" to be used in page headers.
\title{FedADB: Class Anchor-Driven Dual-Branch Federated Learning for Mitigating Forgetting}

%%
%% The "author" command and its associated commands are used to define
%% the authors and their affiliations.
%% Of note is the shared affiliation of the first two authors, and the
%% "authornote" and "authornotemark" commands
%% used to denote shared contribution to the research.

\author{Zhenyan Liu}
\affiliation{%
  \institution{State Key Laboratory of Networking and Switching Technology, Beijing University of Posts and Telecommunications}
  \city{Beijing}
  \country{China}}
\email{lzhy@bupt.edu.cn}

\author{Hua Zhang}
\correspondingauthor
\affiliation{%
  \institution{State Key Laboratory of Networking and Switching Technology, Beijing University of Posts and Telecommunications}
  \city{Beijing}
  \country{China}}
\additionalaffiliation{%
  \institution{National Engineering Research Center of Disaster Backup and Recovery, Beijing University of Posts and Telecommunications}
  \city{Beijing}
  \country{China}}
\email{zhanghua\_288@bupt.edu.cn}

\author{Haoran Gao}
\affiliation{%
  \institution{JIUTIAN Research, China Mobile}
  \city{Beijing}
  \country{China}}
\email{gaohaoranit@chinamobile.com}

\author{Qi Li}
\affiliation{%
  \institution{School of Cyberspace Security, Beijing University of Posts and Telecommunications}
  \city{Beijing}
  \country{China}}
\email{liqi2001@bupt.edu.cn}

\author{Hongliang Zhu}
\affiliation{%
  \institution{School of Cyberspace Security, Beijing University of Posts and Telecommunications}
  \city{Beijing}
  \country{China}}
\email{zhuhongliang@bupt.edu.cn}

\author{Huiyu Zhou}
\affiliation{%
  \institution{University of Leicester}
  \city{Leicester}
  \country{United Kingdom}}
\email{hz143@leicester.ac.uk}

\author{Zongliang Shen}
\affiliation{%
  \institution{State Key Laboratory of Networking and Switching Technology, Beijing University of Posts and Telecommunications}
  \city{Beijing}
  \country{China}}
\email{shenzongliang@bupt.edu.cn}

\author{Yanxin Xu}
\affiliation{%
  \institution{State Key Laboratory of Networking and Switching Technology, Beijing University of Posts and Telecommunications}
  \city{Beijing}
  \country{China}}
\email{yanxinxu@bupt.edu.cn}

\author{Jiahui Wang}
\affiliation{%
  \institution{State Key Laboratory of Networking and Switching Technology, Beijing University of Posts and Telecommunications}
  \city{Beijing}
  \country{China}}
\email{jiahuiwang@bupt.edu.cn}

%%
%% By default, the full list of authors will be used in the page
%% headers. Often, this list is too long, and will overlap
%% other information printed in the page headers. This command allows
%% the author to define a more concise list
%% of authors' names for this purpose.
\renewcommand{\shortauthors}{Zhanyan Liu et al.}

%%
%% The abstract is a short summary of the work to be presented in the
%% article.
\begin{abstract}
Multimodal data collected by heterogeneous devices are used for collaborative training, where federated learning (FL) serves as a key paradigm for effective distributed modeling with data privacy preservation. However, local training suffers from the forgetting of previously learned global knowledge under cross-client data heterogeneity, which leads to significant declines in both performance and convergence speed. Most previous studies rely on global alignment strategies to retain global knowledge, which hinder local optimization and lead to inadequate supervision of missing classes. Some studies introduce proxy datasets to supplement supervision for missing classes. However, it remains a challenge to balance class-wise global consistency and local optimization objectives without proxy datasets. In this work, we propose FedADB, a Class Anchor–Driven Dual-Branch FL framework. Specifically, the server generates class anchors optimized in a differentiable input space, which are shared across clients. These class anchors serve as global references that provide supervision for missing classes during local training. A dual-branch collaborative training mechanism is designed for clients. In this mechanism, the anchor-based global branch focuses on learning with global consistency, achieving global knowledge alignment by class-anchor balanced sampling. The local calibration branch focuses on learning discriminative local features, mitigating the degradation of local representations caused by excessive global alignment. Extensive experiments across multiple medical and natural datasets demonstrate that FedADB achieves significant improvements in both accuracy and convergence speed.
\end{abstract}

%%
%% The code below is generated by the tool at http://dl.acm.org/ccs.cfm.
%% Please copy and paste the code instead of the example below.
%%


%%
%% Keywords. The author(s) should pick words that accurately describe
%% the work being presented. Separate the keywords with commas.
%% A "teaser" image appears between the author and affiliation
%% information and the body of the document, and typically spans the
%% page.

%% This command processes the author and affiliation and title
%% information and builds the first part of the formatted document.
\maketitle

\section{Introduction}
The widespread deployment of digital terminals and sensing devices has driven an explosive growth of multimedia data generated by distributed sources such as medical institutions, mobile terminals, and surveillance systems\cite{jani12023leveraging,reddi2025generative,wang2025empowering}. These data have substantially advanced multimedia intelligence, enabling applications such as autonomous driving, fraud detection, and clinical decision support. However, due to legal, ethical, and institutional constraints, these data cannot be used for centralized training. Federated learning (FL) enables clients to collaboratively train models without sharing raw data\cite{hu2024aggregation,cai2024towards,liu2023cross,qi2022clustering}. This framework significantly improves the utility of isolated data and facilitates the collaborative optimization of neural network models \cite{meng2024improving,wang2024fednlr,cai2024lg,wang2023dafkd}. However, in real-world applications, data across devices are typically Non-Independent and Identically Distributed (non-IID)\cite{yan2025simple,yang2021achieving,reisizadeh2020robust,hsu2019measuring}. Recent studies reveal that under such statistical heterogeneous data distributions, local training on client data leads to a divergence between local and global optimization objectives \cite{li2020federated,shi2025fedawa}. This divergence leads to overfitting to local data distributions while forgetting knowledge acquired from previous training rounds. This phenomenon is known as Catastrophic Forgetting (CF) \cite{mccloskey1989catastrophic}. Consequently, the global model suffers from significant performance degradation and a notable decline in convergence speed \cite{lee2024fedsol,huang2022learn,guo2023fedbr,shoham2019overcoming}.\\
\indent To alleviate issues of CF, existing approaches can roughly be categorized into two main types. The first category typically guides local optimization through gradient or objective function adjustments derived from the global model \cite{shoham2019overcoming,karimireddy2020scaffold,acar2021federated,luo2023gradma}. This technique ensures that local updates maintain consistency with the global optimization direction. The second category usually employs knowledge distillation (KD) to preserve global knowledge \cite{huang2022learn,lee2022preservation,aljahdali2025flashback,lu2023federated}. This paradigm leverages the global model's outputs to guide local training, thereby preserving knowledge acquired from previous tasks. Inherently, both categories employ alignment strategies that reference the global model to calibrate local updates. Although these methods can mitigate CF through knowledge transfer and variance control, they are prone to local-global optimization conflicts and may suppress local learning due to excessive alignment \cite{lee2024fedsol,song2024feddistill}. More importantly, these methods struggle to provide effective supervision for classes missing in clients' local data. Specifically, the feature representations and decision boundaries for missing classes remain unconstrained during local optimization. Consequently, the global model exhibits severe performance degradation on these classes. To address this issue, recent studies provide proxy datasets to clients for knowledge compensation \cite{chen2023towards,guo2023fedbr}. These studies introduce explicit supervision for clients' missing classes, which helps maintain accuracy for those classes. However, proxy datasets are often unavailable in privacy-sensitive domains \cite{zhu2023weak,wu2024exploring,zhu2021data}, and may be limited by cross-domain distribution shifts.\\
\indent To address these challenges, this paper proposes Class Anchor-Driven Dual-Branch Federated Learning (FedADB), a framework that includes two main components: a Class Anchor Generation (CAG) method and a Dual-Branch Cooperative Training (DBCT) mechanism. Specifically, the CAG method generates shared class anchors for each class via differentiable optimization in the input space. These anchors are incorporated into local training as global references to supply supervision for missing classes. During local update, the DBCT mechanism balances global and local optimization objectives via two complementary branches: The Anchor-based Global Branch (AGB) maintains global consistency by leveraging class anchors and balanced sampling, thereby preserving and propagating global knowledge. The Local Calibration Branch (LCB) captures each client’s distinctive data distribution by optimizing directly on the client-specific local dataset. This design avoids the degradation of local representation capability due to excessive alignment with the global objective. Under the coordinated guidance of both branches, a shared feature extractor learns representations that are simultaneously globally generalizable and locally adaptive.\\
\indent Experiments were conducted on multiple natural image classification datasets and medical image classification datasets, including extensive evaluation includes performance comparisons, ablation studies, analysis by case studies. The results validate that FedADB effectively mitigates catastrophic forgetting under non-IID data settings. Our main contributions are as follows:
\begin{itemize}
  \item We propose a class anchor generation method CAG. Unlike prior approaches that rely on external data or auxiliary priors, CAG constructs anchors solely from the global model, thereby avoiding additional privacy-leakage risks. Moreover, each class requires only a single anchor sample to achieve lightweight yet effective global knowledge preservation.
 \item We propose a dual-branch cooperative training mechanism DBCT. Unlike existing FL training schemes, DBCT employs differentiated training strategies with branch-specific optimization objectives. One branch preserves global knowledge through class anchor constraints and class-balanced sampling, while the other focuses on local data adaptation. This design enables the model to retain global knowledge while effectively learning local feature representations.
 \item Extensive experiments demonstrate that FedADB effectively mitigates catastrophic forgetting while improving the smoo\-thness of the global model. FedADB achieves superior or competitive performance against state-of-the-art (SOTA) methods in both accuracy and convergence speed.
\end{itemize}

\section{Related Work}
\subsection{Heterogeneity in Federated Learning}
FL enables clients to collaboratively train models without sharing their raw data. However, the mainstream algorithm FedAvg \cite{mcmahan2017communication} suffers substantial performance degradation under severe data heterogeneity, which has motivated numerous refinement strategies \cite{li2022federated,zhu2021federated}. The first category mitigates client drift by constraining or correcting local updates. Representative methods include FedProx \cite{li2020federated}, SCAFFOLD \cite{karimireddy2020scaffold}, MOON \cite{li2021model}, and FedDyn \cite{acar2021federated}. Another category focuses on server-side aggregation, such as FedAvgM \cite{hsu2019measuring}, which incorporates momentum into global model updates. Additionally, personalized FL instead learns client-specific models adapted to individual data distributions \cite{t2020personalized,tan2022towards}. Representative strategies include fine-tuning the global model \cite{fallah2020personalized}, introducing personalized projection heads \cite{collins2021exploiting}, and designing adaptive aggregation schemes \cite{huang2021personalized,zhang2023fedala}. Although these methods alleviate data heterogeneity, their generalizability and robustness remain limited. In particular, some methods do not consistently outperform FedAvg across heterogeneous settings, as shown in our experiments. This underscores the necessity for more robust FL frameworks capable of fundamentally resolving data heterogeneity

\subsection{Forgetting in Local Learning}
CF primarily results from distribution shifts during local training in FL, which cause clients to forget out-of-distribution knowledge and degrade global performance\cite{huang2022learn,guo2023fedbr,legate2023re}. Unlike the sequential data problem in continual learning, this paper focuses on mitigating CF across spatially distributed data. FedCurv \cite{shoham2019overcoming} was among the earliest studies to identify CF in FL. Subsequent research has predominantly followed two directions. The first employs optimization-based guidance to mitigate knowledge forgetting during local training \cite{lee2024fedsol,luo2023gradma}. The second category employs KD. For instance, FedNTD \cite{lee2022preservation} preserves knowledge of non-target classes by masking target-class logits, whereas FedLMD \cite{lu2023federated} transfers knowledge for minority-class by masking majority-label predictions. However, local updates constrained by global knowledge tend to suppress local learning and offer inadequate supervision for the missing classes. To address this, several methods introduce proxy data to preserve missing-class knowledge. Flashback \cite{aljahdali2025flashback} leverages public datasets to fuse knowledge across rounds during aggregation. FedBR \cite{guo2023fedbr} and FedKA \cite{chen2023towards} reduce local learning bias with globally shared proxy datasets. Since proxy datasets are often unavailable, we generate class anchors by optimizing noise via the global model. We further develop a dual-branch collaborative training framework that jointly preserves global knowledge and facilitates local learning.

\section{Preliminaries}

Consider a FL system with a central server and $K$ clients. Client $k\in[K]$ holds a local dataset $\mathcal{D}_k=\{(\mathbf{x}_i,y_i)\}_{i=1}^{|\mathcal{D}_k|}$ with a label space $\mathcal{C}_k \subseteq \mathcal{C}$, where $\mathcal{C}$ denotes the global label space. FL aims to learn a global model $\mathbf{w}_G$ that generalizes across the global data distribution. The global optimization objective is formulated as
\begin{equation}
\min_{\mathbf{w}_G}\mathcal{L}(\mathbf{w}_G)=\sum_{k=1}^{K}\frac{\left|\mathcal{D}_k\right|}{\sum_{k\in\left[K\right]}\left|\mathcal{D}_k\right|}\mathcal{L}_k(\mathbf{w}_G),
\end{equation}
where $\mathcal{L}_k$ is the local objective of client $k$.
At communication round $t$, the server broadcasts $\mathbf{w}_G^t$ to the participating client set $\mathcal{K}^t \subseteq [K]$. For  client $k$, the local objective is defined as
\begin{equation}
\mathcal{L}_k(\mathbf{w}_G^t)=\mathbb{E}_{(\mathbf{x},y)\sim\mathcal{D}_k}\left[\ell_k\bigl(f(\mathbf{x};\mathbf{w}_G^t),y\bigr)\right],
\end{equation}
where $f(\mathbf{x};\mathbf{w})$ denotes the model prediction, and $\ell$ is the sample-wise loss. After local training, the server aggregates the updated local models with weights $\sigma_k$ to obtain the updated global model $\mathbf{w}_G^{t+1}=\sum_{k\in\mathcal{K}^t}\sigma_k\mathbf{w}_k^{t+1}$. This procedure is repeated for $T$  rounds.

% 图片环境
\begin{figure*}[ht]
    \centering
    \includegraphics[width=0.82\linewidth]{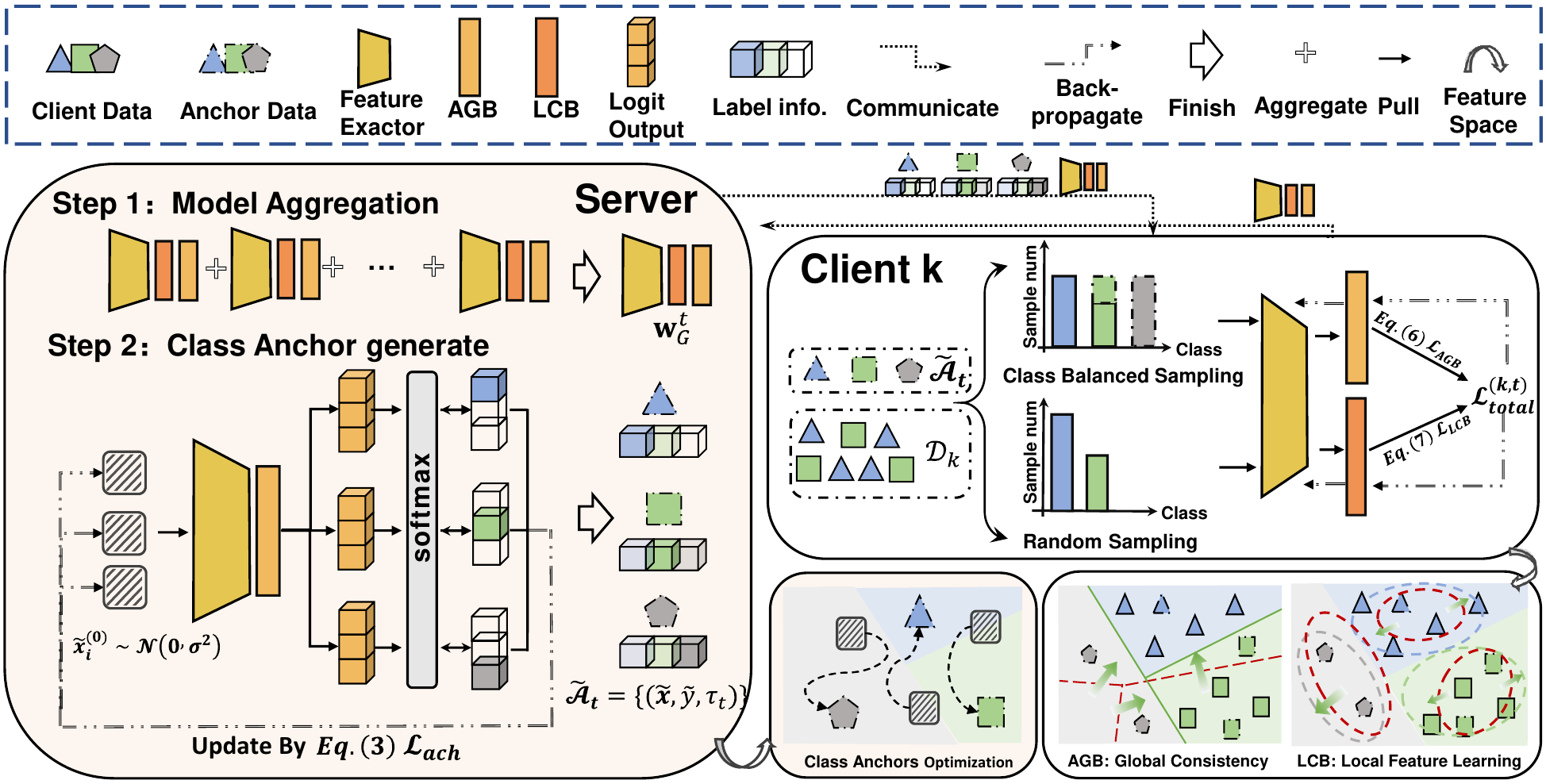}  
    \caption{Overview of FedADB. \Circled{1} After aggregation, the server generates class anchors and broadcasts them to the clients. \Circled{2} Each client performs dual-branch collaborative training: the AGB constructs class-balanced mixed batches from class anchors and local data, while the LCB is optimized using only local data. The shared feature extractor is jointly optimized by both branches. The updated parameters are then uploaded to the server.}
        \Description{See caption.}
    \label{fig:Frame}
\end{figure*}
\section{Methodology}
\label{sec:Method}
\subsection{Overview of FedADB}
FL training often suffers from inadequate supervision for missing classes and an inherent conflict between global consistency and local adaptation, both of which jointly aggravate catastrophic forgetting. FedADB addresses both issues by supplying class-level global supervision without forcing
local optimization to over-align with the global objective. The overall framework of FedADB is shown in Fig.~\ref{fig:Frame}. Specifically, \Circled{1} the server first generates class anchors from the aggregated model and broadcasts them as shared supervisory signals. \Circled{2} Each client performs dual-branch collaborative training, comprising an Anchor-guided Global Branch (AGB), which preserves globally consistent decision boundaries through anchor-guided optimization, whereas a Local Calibration Branch (LCB) leverages local data to maintain client-specific discriminative capability. The two branches share a feature extractor and are coupled through an adaptive loss
weight, yielding a more effective trade-off between global alignment and local adaptation. 
\subsection{Class Anchor Generation}\label{sec:CAG} 
To mitigate the performance degradation caused by insufficient supervision of missing classes, we supplement the supervisory signal by introducing class anchors. After obtaining the global model, the server initializes $\widetilde{\mathbf{x}}_i^{(0)}\sim\mathcal{N}(\mathbf{0},\sigma^2\mathbf{I})$ for target class $c_i\in\mathcal{C}$, then optimizes them via the feature extractor $\phi(\cdot;\boldsymbol{\theta}_\phi^t)$ and AGB classifier $\psi_A(\cdot;\boldsymbol{\theta}_A^t)$. The optimization objective is defined as
\begin{equation}
\begin{aligned}
\mathcal{L}_{\mathrm{ach}}(\widetilde{\mathbf{x}}_i;c_i)
={}&-\log\!\left[
\psi_A\!\left(
\phi(\widetilde{\mathbf{x}}_i;\boldsymbol{\theta}_\phi^t);
\boldsymbol{\theta}_A^t
\right)
\right]_{c_i} +\lambda\|\widetilde{\mathbf{x}}_i\|_2^2,
\end{aligned}
\end{equation}
where the first term is the negative log-likelihood (NLL) loss, which explicitly drives $\widetilde{\mathbf{x}}_i$ toward class $c_i$ under the current global model, while the second term is an $L_2$ regularizer with coefficient $\lambda$ that penalizes extreme values during optimization. The class-anchor update at iteration $j$ is given by
\begin{equation}
\widetilde{\mathbf{x}}_i^{(j+1)}
=\Pi_{\mathcal{X}}\!\left(
\widetilde{\mathbf{x}}_i^{(j)}
-\eta\nabla_{\widetilde{\mathbf{x}}_i^{(j)}}
\mathcal{L}_{\mathrm{ach}}(\widetilde{\mathbf{x}}_i^{(j)};c_i)
\right),
\end{equation}
where $\eta$ is the learning rate. Starting from the Gaussian initialization, we iteratively update $\widetilde{\mathbf{x}}_i$ to progressively increase the target class posterior. $\Pi_{\mathcal{X}}$ projects each update onto the valid input domain $\mathcal{X}$. This improves numerical stability and keeps anchor generation consistent with the data space. After $I$ optimization iterations, we obtain a total of $N$ class anchors. For anchor $i$, we define its target-class confidence as ${\widetilde{y}}_i$ and quantify the overall anchor quality by
\begin{equation}
\tau_t=\frac{1}{N}\sum_{i=1}^{N}[\psi_A(\phi({\widetilde{\mathbf{x}}}_i;\boldsymbol{\theta}^t_\phi);\boldsymbol{\theta}^t_A)]_{c_i},
\end{equation}
The server broadcasts the anchor set $\widetilde{\mathcal{A}}_t=\{({\widetilde{\mathbf{x}}},{\widetilde{y}},\tau_t)\}$ to all participating clients, providing a lightweight yet consistent global supervisory signal. Because the global model changes after every aggregation, the server regenerates the anchors at each round to keep the supervisory signal synchronized with the current model.
\subsection{Dual-Branch Collaborative Training}
To mitigate the conflict between global consistency and local adaptability, we propose dual-branch collaborative training (DBCT). The model comprises a shared feature extractor  $\phi(\cdot;\boldsymbol{\theta}_\phi)$ and two complementary classification branches: the branch AGB for global alignment and the branch LCB for local representation learning. Gradients from both branches jointly optimize the shared extractor $\boldsymbol{\theta}_\phi$, while each branch follows a distinct  optimization strategy.

\textbf{Anchor-guided Global Branch (AGB).} For the input $\mathbf{x}$, the AGB employs a classifier $\psi_A(\cdot;\boldsymbol{\theta}_A)$ to produce predictions
$f_A(\mathbf{x})=\psi_A(\phi(\mathbf{x};\boldsymbol{\theta}_\phi);\boldsymbol{\theta}_A)$. For client $k$, we construct a class-balanced mixed batch $\mathcal{B}_{k}^{A}$ from local samples and anchors in $\widetilde{\mathcal{A}}_t$, ensuring that every class can contribute supervisory signals even when it is absent from $\mathcal{D}_k$. Let $\mathcal{H}(\cdot,\cdot)$ denotes the cross-entropy loss. The AGB loss is
\begin{equation}
\mathcal{L}_{AGB}=\mathbb{E}_{(\mathbf{x},y)\sim\mathcal{B}_k^A}\mathcal{H}(y,f_A(\mathbf{x})),
\end{equation}
 By minimizing $\mathcal{L}_{AGB}$, class anchors provide stable gradient constraints across the global label space, thereby preserving global knowledge and mitigating forgetting during local updates.

\textbf{Local Calibration Branch (LCB).} For the input $\mathbf{x}$, the LCB employs a classifier $\psi_L(\cdot;\boldsymbol{\theta}_L)$ and outputs
$f_L(\mathbf{x})=\psi_L(\phi(\mathbf{x};\boldsymbol{\theta}_\phi);\boldsymbol{\theta}_L)$. Unlike the AGB, this branch is optimized solely on the local data $\mathcal{D}_k$ with a batch size aligned to $\mathcal{B}_k^A$, so that it can captures client-dependent class priors and feature statistics. As a result, the LCB preserves local discriminative structure and compensates for representation bias introduced by strict global alignment. Its loss is
\begin{equation}
\mathcal{L}_{LCB}=\mathbb{E}_{(\mathbf{x},y)\sim\mathcal{D}_k}\mathcal{H}(y,f_L(\mathbf{x})),
\end{equation}
Let $r={t}/{T}$ denote the FL progress. We define the LCB weight for client $k$ as $\beta_k(r)=r^{\gamma_k\cdot\tau_t}$, where $\gamma_k=1-{|\mathcal{C}_k|}/{|\mathcal{C}|}$ represents the class-missing rate for client $k$, and $\tau_t$ measures the quality of the anchors. When $\gamma_k$ or $\tau_t$ is high, $\beta^t_k(r)$ is attenuated to emphasize global consistency. Otherwise, $\beta^t_k(r)$ increases more rapidly to strengthen local learning. The total loss for client $k$ is given by

\begin{equation}
\mathcal{L}_{total}^{(k,t)}=\mathcal{L}_{AGB}+\beta^t_k(r)\cdot\mathcal{L}_{LCB},
\end{equation}
Client $k$ minimizes $\mathcal{L}_{\mathrm{total}}^{k,t}$ with respect to $\{\boldsymbol{\theta}_\phi,\boldsymbol{\theta}_A,\boldsymbol{\theta}_L\}$ and returns the updated parameters to the server for aggregation. Through the decoupling of knowledge preservation and local learning, DBCT enables effective adaptation to local data while maintaining global knowledge, thereby alleviating catastrophic forgetting.

\newcommand{\tabstdsize}{\scriptsize}

\begin{table*}[t]
\caption{Comparison of test accuracy@1 (\%) among the baselines and FedADB on different datasets under cross-device and cross-silo scenarios. We set $\alpha \in \{0.05, 0.1, 0.2, 0.5\}$ for CIFAR-10 and $\alpha=0.1$ for the other datasets. Bold indicates the best result.}
\label{tab:main_results}
\centering

{
\small
\setlength{\tabcolsep}{4pt}
\renewcommand{\arraystretch}{0.85}

\begin{tabular}{
c@{\hspace{4pt}}
c@{\hspace{4pt}}
c@{\hspace{4pt}}
c@{\hspace{4pt}}
c@{\hspace{4pt}}
c@{\hspace{4pt}}
c@{\hspace{4pt}}
c@{\hspace{4pt}}
c@{\hspace{4pt}}
c@{\hspace{4pt}}
c@{\hspace{4pt}}
c
}

\toprule

% ==================== Cross-Device ====================

\multirow[c]{2}{*}{\textbf{Method}}
& \multirow[c]{2}{*}{\textbf{SVHN}}
& \multirow[c]{2}{*}{\textbf{CINIC-10}}
& \multicolumn{4}{c}{\textbf{CIFAR-10}}
& \multirow[c]{2}{*}{\textbf{CIFAR-100}}
& \multirow[c]{2}{*}{\makecell[c]{\textbf{OrganA-}\\\textbf{MNIST}}}
& \multirow[c]{2}{*}{\makecell[c]{\textbf{Path-}\\\textbf{MNIST}}}
& \multirow[c]{2}{*}{\makecell[c]{\textbf{Tissue-}\\\textbf{MNIST}}}
& \multirow[c]{2}{*}{\textbf{Average}}
\\
\cmidrule(lr){4-7}

&
&
&
$\alpha=0.05$
&
$\alpha=0.1$
&
$\alpha=0.2$
&
$\alpha=0.5$
&
&
&
&
&
\\
\midrule
\multicolumn{12}{c}{\textbf{Cross-Device Scenario ($K=100,\rho=0.05$)}} \\
\midrule

FedAvg\cite{mcmahan2017communication}
& 84.56{\tabstdsize (±1.57)}
& 43.55{\tabstdsize (±1.96)}
& 32.32{\tabstdsize (±1.40)}
& 48.76{\tabstdsize (±1.14)}
& 57.66{\tabstdsize (±1.43)}
& 66.76{\tabstdsize (±0.81)}
& 47.54{\tabstdsize (±0.31)}
& 26.68{\tabstdsize (±2.78)}
& 69.01{\tabstdsize (±3.19)}
& 21.84{\tabstdsize (±5.07)}
& 49.87{\tabstdsize (±1.97)}
\\

FedDyn\cite{acar2021federated}
& 83.49{\tabstdsize (±4.28)}
& 44.69{\tabstdsize (±1.00)}
& 28.38{\tabstdsize (±0.10)}
& 42.62{\tabstdsize (±2.66)}
& \textbf{66.91}{\tabstdsize (±1.91)}
& 69.38{\tabstdsize (±0.52)}
& 42.87{\tabstdsize (±0.89)}
& 66.10{\tabstdsize (±3.61)}
& 74.73{\tabstdsize (±1.59)}
& 32.11{\tabstdsize (±0.68)}
& 55.13{\tabstdsize (±1.72)}
\\

FedASAM\cite{caldarola2022improving}
& 73.86{\tabstdsize (±3.31)}
& 46.38{\tabstdsize (±1.44)}
& 30.80{\tabstdsize (±2.04)}
& 47.19{\tabstdsize (±2.33)}
& 62.63{\tabstdsize (±0.39)}
& 69.64{\tabstdsize (±0.28)}
& 46.79{\tabstdsize (±0.85)}
& 58.31{\tabstdsize (±1.47)}
& 71.85{\tabstdsize (±1.30)}
& 28.23{\tabstdsize (±3.51)}
& 53.57{\tabstdsize (±1.69)}
\\

FedNTD\cite{lee2022preservation}
& 89.23{\tabstdsize (±0.44)}
& 41.78{\tabstdsize (±0.75)}
& 39.64{\tabstdsize (±2.17)}
& 54.03{\tabstdsize (±1.00)}
& 59.53{\tabstdsize (±0.52)}
& 70.86{\tabstdsize (±0.40)}
& 46.19{\tabstdsize (±1.23)}
& 64.61{\tabstdsize (±1.03)}
& 77.31{\tabstdsize (±1.48)}
& 15.65{\tabstdsize (±2.93)}
& 55.88{\tabstdsize (±1.20)}
\\

FedGELA\cite{fan2023federated}
& 89.25{\tabstdsize (±0.46)}
& \textbf{48.01}{\tabstdsize (±0.84)}
& 41.86{\tabstdsize (±1.53)}
& 56.84{\tabstdsize (±0.72)}
& 61.27{\tabstdsize (±4.05)}
& 66.42{\tabstdsize (±0.47)}
& 35.45{\tabstdsize (±1.28)}
& 45.75{\tabstdsize (±1.72)}
& 73.57{\tabstdsize (±1.99)}
& 19.26{\tabstdsize (±3.43)}
& 53.77{\tabstdsize (±1.65)}
\\

FedEXP\cite{jhunjhunwala2023fedexp}
& 69.10{\tabstdsize (±0.81)}
& 41.05{\tabstdsize (±0.67)}
& 40.64{\tabstdsize (±0.81)}
& 49.73{\tabstdsize (±0.16)}
& 55.56{\tabstdsize (±1.06)}
& 65.74{\tabstdsize (±0.05)}
& 46.04{\tabstdsize (±0.52)}
& 73.49{\tabstdsize (±1.79)}
& 73.55{\tabstdsize (±0.81)}
& 30.29{\tabstdsize (±0.80)}
& 54.52{\tabstdsize (±0.75)}
\\

FedLMD\cite{lu2023federated}
& 89.80{\tabstdsize (±0.14)}
& 42.99{\tabstdsize (±0.87)}
& 42.47{\tabstdsize (±0.48)}
& 55.87{\tabstdsize (±0.21)}
& 59.68{\tabstdsize (±0.92)}
& 70.80{\tabstdsize (±0.35)}
& 38.50{\tabstdsize (±1.88)}
& 65.13{\tabstdsize (±2.05)}
& 74.75{\tabstdsize (±1.40)}
& 14.54{\tabstdsize (±0.68)}
& 55.45{\tabstdsize (±0.90)}
\\

FedDecorr\cite{shi2023towards}
& 88.99{\tabstdsize (±1.16)}
& 41.51{\tabstdsize (±1.15)}
& 37.77{\tabstdsize (±1.92)}
& 55.15{\tabstdsize (±0.41)}
& 58.11{\tabstdsize (±0.52)}
& 66.88{\tabstdsize (±0.37)}
& 48.15{\tabstdsize (±0.90)}
& 71.39{\tabstdsize (±1.49)}
& 67.18{\tabstdsize (±1.15)}
& 35.15{\tabstdsize (±0.61)}
& 57.03{\tabstdsize (±0.97)}
\\

FedACG\cite{kim2024communication}
& 83.64{\tabstdsize (±0.65)}
& 43.98{\tabstdsize (±0.68)}
& 30.61{\tabstdsize (±1.32)}
& 51.18{\tabstdsize (±1.55)}
& 56.76{\tabstdsize (±1.66)}
& 66.83{\tabstdsize (±0.41)}
& 47.49{\tabstdsize (±0.49)}
& 65.90{\tabstdsize (±3.37)}
& 66.97{\tabstdsize (±2.10)}
& 23.70{\tabstdsize (±2.80)}
& 53.71{\tabstdsize (±1.50)}
\\

FedSOL\cite{lee2024fedsol}
& 87.14{\tabstdsize (±0.33)}
& 41.60{\tabstdsize (±0.18)}
& 42.54{\tabstdsize (±0.96)}
& 50.38{\tabstdsize (±0.23)}
& 56.32{\tabstdsize (±1.17)}
& 65.61{\tabstdsize (±0.15)}
& 38.33{\tabstdsize (±0.18)}
& 67.54{\tabstdsize (±2.02)}
& 72.23{\tabstdsize (±2.74)}
& 33.03{\tabstdsize (±0.81)}
& 55.47{\tabstdsize (±0.88)}
\\

\midrule

\textbf{FedADB (Ours)}
& \textbf{90.06}{\tabstdsize (±0.40)}
& 47.67{\tabstdsize (±0.69)}
& \textbf{50.06}{\tabstdsize (±0.61)}
& \textbf{59.21}{\tabstdsize (±0.47)}
& 64.55{\tabstdsize (±0.56)}
& \textbf{70.97}{\tabstdsize (±0.29)}
& \textbf{57.70}{\tabstdsize (±0.31)}
& \textbf{78.04}{\tabstdsize (±1.66)}
& \textbf{79.08}{\tabstdsize (±0.53)}
& \textbf{40.75}{\tabstdsize (±1.11)}
& \textbf{63.81}{\tabstdsize (±0.66)}
\\\hline

% ==================== Cross-Silo ====================
\midrule

\multicolumn{12}{c}{
    \textbf{Cross-Silo Scenario ($K=20,\rho=0.2$)}
} \\
\midrule

FedAvg\cite{mcmahan2017communication}
& 91.90{\tabstdsize (±0.15)}
& 35.02{\tabstdsize (±0.06)}
& 31.67{\tabstdsize (±1.17)}
& 55.96{\tabstdsize (±1.49)}
& 66.17{\tabstdsize (±0.43)}
& 72.96{\tabstdsize (±0.49)}
& 63.78{\tabstdsize (±0.16)}
& 67.44{\tabstdsize (±0.17)}
& 75.26{\tabstdsize (±1.38)}
& 39.81{\tabstdsize (±0.50)}
& 60.00{\tabstdsize (±0.60)}
\\

FedDyn\cite{acar2021federated}
& 89.75{\tabstdsize (±0.13)}
& 37.22{\tabstdsize (±0.33)}
& 31.95{\tabstdsize (±0.88)}
& 58.59{\tabstdsize (±0.95)}
& 67.53{\tabstdsize (±0.44)}
& 72.67{\tabstdsize (±0.22)}
& 60.40{\tabstdsize (±0.22)}
& 67.36{\tabstdsize (±0.37)}
& 76.11{\tabstdsize (±1.44)}
& 38.67{\tabstdsize (±0.92)}
& 60.03{\tabstdsize (±0.59)}
\\

FedASAM\cite{caldarola2022improving}
& \textbf{93.00}{\tabstdsize (±0.36)}
& 38.79{\tabstdsize (±2.46)}
& 35.29{\tabstdsize (±1.75)}
& 59.74{\tabstdsize (±0.70)}
& 69.81{\tabstdsize (±0.50)}
& 75.45{\tabstdsize (±0.32)}
& 65.15{\tabstdsize (±0.77)}
& 68.90{\tabstdsize (±0.31)}
& 78.27{\tabstdsize (±0.69)}
& 39.80{\tabstdsize (±1.53)}
& 62.42{\tabstdsize (±0.94)}
\\

FedNTD\cite{lee2022preservation}
& 91.32{\tabstdsize (±0.12)}
& 36.69{\tabstdsize (±0.35)}
& 34.14{\tabstdsize (±1.37)}
& 60.02{\tabstdsize (±1.54)}
& 71.32{\tabstdsize (±0.56)}
& \textbf{76.51}{\tabstdsize (±0.48)}
& 63.35{\tabstdsize (±0.08)}
& 67.90{\tabstdsize (±1.54)}
& 73.98{\tabstdsize (±1.77)}
& 36.00{\tabstdsize (±0.51)}
& 61.12{\tabstdsize (±0.83)}
\\

FedGELA\cite{fan2023federated}
& 92.30{\tabstdsize (±0.24)}
& 33.50{\tabstdsize (±0.18)}
& 32.17{\tabstdsize (±1.21)}
& 59.44{\tabstdsize (±1.56)}
& 64.96{\tabstdsize (±0.78)}
& 70.27{\tabstdsize (±1.00)}
& 60.43{\tabstdsize (±0.30)}
& 59.22{\tabstdsize (±0.43)}
& 70.06{\tabstdsize (±1.65)}
& 33.95{\tabstdsize (±4.05)}
& 57.63{\tabstdsize (±1.14)}
\\

FedEXP\cite{jhunjhunwala2023fedexp}
& 92.06{\tabstdsize (±0.20)}
& 36.35{\tabstdsize (±0.18)}
& 33.28{\tabstdsize (±0.37)}
& 58.38{\tabstdsize (±1.44)}
& 66.49{\tabstdsize (±0.10)}
& 71.79{\tabstdsize (±0.09)}
& 63.69{\tabstdsize (±0.31)}
& 70.36{\tabstdsize (±1.17)}
& 75.19{\tabstdsize (±0.36)}
& 41.24{\tabstdsize (±0.82)}
& 60.88{\tabstdsize (±0.50)}
\\

FedLMD\cite{lu2023federated}
& 91.78{\tabstdsize (±0.45)}
& 37.19{\tabstdsize (±0.22)}
& 34.25{\tabstdsize (±0.60)}
& 61.89{\tabstdsize (±0.06)}
& 71.67{\tabstdsize (±0.46)}
& 75.89{\tabstdsize (±0.23)}
& 53.95{\tabstdsize (±0.87)}
& 68.73{\tabstdsize (±0.18)}
& 74.93{\tabstdsize (±1.51)}
& 40.05{\tabstdsize (±1.19)}
& 61.03{\tabstdsize (±0.58)}
\\

FedDecorr\cite{shi2023towards}
& 91.96{\tabstdsize (±0.10)}
& 36.06{\tabstdsize (±0.61)}
& 32.49{\tabstdsize (±1.31)}
& 58.42{\tabstdsize (±0.53)}
& 65.44{\tabstdsize (±1.11)}
& 72.35{\tabstdsize (±0.23)}
& 62.85{\tabstdsize (±0.46)}
& 70.99{\tabstdsize (±0.70)}
& 74.83{\tabstdsize (±1.21)}
& 39.56{\tabstdsize (±0.55)}
& 60.50{\tabstdsize (±0.68)}
\\

FedACG\cite{kim2024communication}
& 91.97{\tabstdsize (±0.12)}
& 35.61{\tabstdsize (±1.22)}
& 31.30{\tabstdsize (±1.38)}
& 56.01{\tabstdsize (±0.50)}
& 65.78{\tabstdsize (±0.66)}
& 72.88{\tabstdsize (±0.43)}
& 63.94{\tabstdsize (±0.14)}
& 67.50{\tabstdsize (±1.54)}
& 75.93{\tabstdsize (±0.72)}
& 37.06{\tabstdsize (±3.58)}
& 59.80{\tabstdsize (±1.03)}
\\

FedSOL\cite{lee2024fedsol}
& 90.98{\tabstdsize (±0.26)}
& 39.35{\tabstdsize (±0.41)}
& 40.13{\tabstdsize (±0.34)}
& 58.08{\tabstdsize (±1.55)}
& 66.49{\tabstdsize (±0.55)}
& 72.75{\tabstdsize (±0.13)}
& 59.67{\tabstdsize (±0.35)}
& 69.41{\tabstdsize (±2.49)}
& 73.08{\tabstdsize (±1.20)}
& 35.86{\tabstdsize (±0.49)}
& 60.58{\tabstdsize (±0.78)}
\\

\midrule

\textbf{FedADB (Ours)}
& 92.36{\tabstdsize (±0.08)}
& \textbf{44.62}{\tabstdsize (±0.44)}
& \textbf{54.02}{\tabstdsize (±0.60)}
& \textbf{67.65}{\tabstdsize (±0.25)}
& \textbf{71.84}{\tabstdsize (±0.33)}
& 75.07{\tabstdsize (±0.48)}
& \textbf{65.57}{\tabstdsize (±0.51)}
& \textbf{78.21}{\tabstdsize (±0.34)}
& \textbf{80.48}{\tabstdsize (±1.27)}
& \textbf{41.45}{\tabstdsize (±0.50)}
& \textbf{67.13}{\tabstdsize (±0.48)}
\\

\bottomrule
\end{tabular}
}
\end{table*}

\begin{table*}[ht]
\centering

\caption{
Communication rounds required to reach $T_{\mathrm{Acc}}$ and the corresponding
relative speedup over FedADB. ``--'' indicates that the method fails to reach
$T_{\mathrm{Acc}}$ and is therefore excluded from the average calculation.
Unless otherwise specified, $\alpha=0.1$.
}
\label{tab:Convergence}

% Relative speedup shown in a smaller font.
\newcommand{\ratio}[1]{\,{\scriptsize\textnormal{(#1$\times$)}}}

\begin{adjustbox}{max width=\textwidth}
\begin{minipage}{\textwidth}
\centering
\small
\setlength{\tabcolsep}{2.4pt}
\renewcommand{\arraystretch}{0.83}

\begin{tabular}{
    c
    c
    c
    *{11}{c}
}
\toprule
\textbf{Dataset}
& \boldmath$\boldsymbol{\alpha}$
& \boldmath$\boldsymbol{T_{\mathrm{Acc}}}$
& \textbf{FedAvg}
& \textbf{FedDyn}
& \textbf{FedASAM}
& \textbf{FedNTD}
& \textbf{FedGELA}
& \textbf{FedEXP}
& \textbf{FedLMD}
& \textbf{FedDecorr}
& \textbf{FedACG}
& \textbf{FedSOL}
& \textbf{FedADB}
\\
\midrule

% =========================
% SVHN
% =========================
\multirow{3}{*}{\textbf{SVHN}}
& \multirow{3}{*}{$0.1$}
& 42\%
& 274\ratio{13.7}
& 96\ratio{4.8}
& 426\ratio{21.3}
& 83\ratio{4.2}
& 80\ratio{4.0}
& 60\ratio{3.0}
& 65\ratio{3.3}
& 159\ratio{8.0}
& 321\ratio{16.1}
& 31\ratio{1.6}
& \textbf{20}\ratio{1.0}
\\

&
& 61\%
& 330\ratio{12.2}
& 157\ratio{5.8}
& 585\ratio{21.7}
& 118\ratio{4.4}
& 106\ratio{3.9}
& 106\ratio{3.9}
& 95\ratio{3.5}
& 184\ratio{6.8}
& 380\ratio{14.1}
& 56\ratio{2.1}
& \textbf{27}\ratio{1.0}
\\

&
& 80\%
& 478\ratio{9.6}
& 272\ratio{5.4}
& --
& 191\ratio{3.8}
& 149\ratio{3.0}
& --
& 147\ratio{2.9}
& 245\ratio{4.9}
& 571\ratio{11.4}
& 81\ratio{1.6}
& \textbf{50}\ratio{1.0}
\\

\midrule

% =========================
% CINIC-10
% =========================
\multirow{3}{*}{\textbf{CINIC-10}}
& \multirow{3}{*}{$0.1$}
& 20\%
& 74\ratio{10.6}
& 38\ratio{5.4}
& 62\ratio{8.9}
& 29\ratio{4.1}
& 93\ratio{13.3}
& 37\ratio{5.3}
& 28\ratio{4.0}
& 38\ratio{5.4}
& 74\ratio{10.6}
& 37\ratio{5.3}
& \textbf{7}\ratio{1.0}
\\

&
& 29\%
& 192\ratio{6.9}
& 112\ratio{4.0}
& 192\ratio{6.9}
& 99\ratio{3.5}
& 241\ratio{8.6}
& 133\ratio{4.8}
& 63\ratio{2.3}
& 191\ratio{6.8}
& 207\ratio{7.4}
& 88\ratio{3.1}
& \textbf{28}\ratio{1.0}
\\

&
& 38\%
& 563\ratio{6.1}
& 215\ratio{2.3}
& 401\ratio{4.3}
& --
& 401\ratio{4.3}
& --
& --
& --
& 562\ratio{6.0}
& 254\ratio{2.7}
& \textbf{93}\ratio{1.0}
\\

\midrule

% =========================
% CIFAR-10, alpha = 0.05
% =========================
\multirow{12}{*}{\textbf{CIFAR-10}}
& \multirow{3}{*}{$0.05$}
& 25\%
& 271\ratio{10.0}
& 356\ratio{13.2}
& 443\ratio{16.4}
& 166\ratio{6.2}
& 217\ratio{8.0}
& 73\ratio{2.7}
& 159\ratio{5.9}
& 199\ratio{7.4}
& 361\ratio{13.4}
& 62\ratio{2.3}
& \textbf{27}\ratio{1.0}
\\

&
& 35\%
& --
& --
& --
& 445\ratio{7.1}
& 356\ratio{5.7}
& 221\ratio{3.5}
& 361\ratio{5.7}
& 445\ratio{7.1}
& --
& 114\ratio{1.8}
& \textbf{63}\ratio{1.0}
\\

&
& 46\%
& --
& --
& --
& --
& --
& --
& --
& --
& --
& --
& \textbf{148}\ratio{1.0}
\\

\cmidrule(lr){2-14}

% =========================
% CIFAR-10, alpha = 0.1
% =========================
&
\multirow{3}{*}{$0.1$}
& 31\%
& 209\ratio{13.1}
& 156\ratio{9.8}
& 225\ratio{14.1}
& 137\ratio{8.6}
& 103\ratio{6.4}
& 79\ratio{4.9}
& 111\ratio{6.9}
& 111\ratio{6.9}
& 208\ratio{13.0}
& 49\ratio{3.1}
& \textbf{16}\ratio{1.0}
\\

&
& 44\%
& 383\ratio{6.6}
& 384\ratio{6.6}
& 388\ratio{6.7}
& 253\ratio{4.4}
& 196\ratio{3.4}
& 204\ratio{3.5}
& 227\ratio{3.9}
& 252\ratio{4.3}
& 381\ratio{6.6}
& 86\ratio{1.5}
& \textbf{58}\ratio{1.0}
\\

&
& 58\%
& --
& --
& --
& 619\ratio{3.9}
& 310\ratio{2.0}
& --
& 519\ratio{3.3}
& 486\ratio{3.1}
& --
& --
& \textbf{157}\ratio{1.0}
\\

\cmidrule(lr){2-14}

% =========================
% CIFAR-10, alpha = 0.2
% =========================
&
\multirow{3}{*}{$0.2$}
& 33\%
& 89\ratio{7.4}
& 65\ratio{5.4}
& 89\ratio{7.4}
& 33\ratio{2.8}
& 64\ratio{5.3}
& 47\ratio{3.9}
& 33\ratio{2.8}
& 68\ratio{5.7}
& 89\ratio{7.4}
& 27\ratio{2.3}
& \textbf{12}\ratio{1.0}
\\

&
& 47\%
& 196\ratio{7.0}
& 125\ratio{4.5}
& 195\ratio{7.0}
& 132\ratio{4.7}
& 149\ratio{5.3}
& 118\ratio{4.2}
& 125\ratio{4.5}
& 175\ratio{6.3}
& 196\ratio{7.0}
& 87\ratio{3.1}
& \textbf{28}\ratio{1.0}
\\

&
& 62\%
& --
& 192\ratio{1.9}
& 415\ratio{4.2}
& 311\ratio{3.1}
& 405\ratio{4.1}
& --
& 310\ratio{3.1}
& 463\ratio{4.7}
& --
& --
& \textbf{99}\ratio{1.0}
\\

\cmidrule(lr){2-14}

% =========================
% CIFAR-10, alpha = 0.5
% =========================
&
\multirow{3}{*}{$0.5$}
& 35\%
& 24\ratio{3.4}
& 20\ratio{2.9}
& 24\ratio{3.4}
& 8\ratio{1.1}
& 20\ratio{2.9}
& 19\ratio{2.7}
& \textbf{7}\ratio{1.0}
& 20\ratio{2.9}
& 24\ratio{3.4}
& 14\ratio{2.0}
& \textbf{7}\ratio{1.0}
\\

&
& 50\%
& 69\ratio{3.5}
& 46\ratio{2.3}
& 68\ratio{3.4}
& 25\ratio{1.3}
& 55\ratio{2.8}
& 53\ratio{2.6}
& 22\ratio{1.1}
& 65\ratio{3.3}
& 69\ratio{3.5}
& 37\ratio{1.9}
& \textbf{20}\ratio{1.0}
\\

&
& 65\%
& 169\ratio{3.1}
& 94\ratio{1.7}
& 150\ratio{2.7}
& 80\ratio{1.5}
& 168\ratio{3.1}
& 132\ratio{2.4}
& 68\ratio{1.2}
& 147\ratio{2.7}
& 168\ratio{3.1}
& 95\ratio{1.7}
& \textbf{55}\ratio{1.0}
\\

\midrule

% =========================
% CIFAR-100
% =========================
\multirow{3}{*}{\textbf{CIFAR-100}}
& \multirow{3}{*}{$0.1$}
& 30\%
& 141\ratio{2.4}
& 110\ratio{1.9}
& 157\ratio{2.7}
& 130\ratio{2.2}
& 244\ratio{4.1}
& 123\ratio{2.1}
& 245\ratio{4.1}
& 129\ratio{2.2}
& 142\ratio{2.4}
& 85\ratio{1.4}
& \textbf{59}\ratio{1.0}
\\

&
& 43\%
& 273\ratio{2.6}
& 234\ratio{2.3}
& 283\ratio{2.7}
& 274\ratio{2.6}
& --
& 237\ratio{2.3}
& 747\ratio{7.2}
& 244\ratio{2.4}
& 272\ratio{2.6}
& 283\ratio{2.7}
& \textbf{104}\ratio{1.0}
\\

&
& 56\%
& --
& --
& --
& --
& --
& --
& --
& --
& --
& --
& \textbf{209}\ratio{1.0}
\\

\midrule

% =========================
% OrganAMNIST
% =========================
\multirow{3}{*}{\textbf{OrganAMNIST}}
& \multirow{3}{*}{$0.1$}
& 36\%
& --
& 65\ratio{10.8}
& 112\ratio{18.7}
& 56\ratio{9.3}
& 134\ratio{22.3}
& 35\ratio{5.8}
& 54\ratio{9.0}
& 38\ratio{6.3}
& 88\ratio{14.7}
& 38\ratio{6.3}
& \textbf{6}\ratio{1.0}
\\

&
& 51\%
& --
& 108\ratio{5.4}
& --
& 115\ratio{5.8}
& --
& 90\ratio{4.5}
& 115\ratio{5.8}
& 90\ratio{4.5}
& 193\ratio{9.7}
& 87\ratio{4.4}
& \textbf{20}\ratio{1.0}
\\

&
& 67\%
& --
& --
& --
& --
& --
& 208\ratio{4.1}
& --
& --
& --
& --
& \textbf{51}\ratio{1.0}
\\

\midrule

% =========================
% PathMNIST
% =========================
\multirow{3}{*}{\textbf{PathMNIST}}
& \multirow{3}{*}{$0.1$}
& 36\%
& 96\ratio{8.7}
& 33\ratio{3.0}
& 83\ratio{7.6}
& 31\ratio{2.8}
& 37\ratio{3.4}
& 34\ratio{3.1}
& 20\ratio{1.8}
& 82\ratio{7.4}
& 96\ratio{8.7}
& 20\ratio{1.8}
& \textbf{11}\ratio{1.0}
\\

&
& 52\%
& 155\ratio{6.0}
& 78\ratio{3.0}
& 155\ratio{6.0}
& 123\ratio{4.7}
& 137\ratio{5.3}
& 82\ratio{3.2}
& 123\ratio{4.7}
& 236\ratio{9.1}
& 235\ratio{9.0}
& 78\ratio{3.0}
& \textbf{26}\ratio{1.0}
\\

&
& 68\%
& --
& 156\ratio{2.1}
& 400\ratio{5.3}
& 264\ratio{3.5}
& 358\ratio{4.7}
& 202\ratio{2.7}
& 268\ratio{3.5}
& --
& --
& 157\ratio{2.1}
& \textbf{76}\ratio{1.0}
\\

\midrule

% =========================
% TissueMNIST
% =========================
\multirow{3}{*}{\textbf{TissueMNIST}}
& \multirow{3}{*}{$0.1$}
& 20\%
& --
& 67\ratio{67.0}
& 262\ratio{262.0}
& --
& --
& 83\ratio{83.0}
& --
& 33\ratio{33.0}
& 411\ratio{411.0}
& 33\ratio{33.0}
& \textbf{1}\ratio{1.0}
\\

&
& 29\%
& --
& --
& --
& --
& --
& --
& --
& 411\ratio{11.4}
& --
& 117\ratio{3.3}
& \textbf{36}\ratio{1.0}
\\

&
& 37\%
& --
& --
& --
& --
& --
& --
& --
& --
& --
& --
& \textbf{222}\ratio{1.0}
\\

\midrule

% =========================
% Average
% =========================
\multicolumn{2}{c}{\textbf{Average}}
& --
& 7.4$\times$
& 7.5$\times$
& 20.6$\times$
& 4.2$\times$
& 5.7$\times$
& 7.2$\times$
& 4.0$\times$
& 6.8$\times$
& 27.7$\times$
& 3.9$\times$
& \textbf{1.0$\times$}
\\

\bottomrule
\end{tabular}
\end{minipage}
\end{adjustbox}
\end{table*}

\section{Experiment}
\subsection{Experimental Setup}

\textbf{Datasets and Models:} We evaluated all methods on four natural image datasets (SVHN \cite{netzer2011reading}, CINIC-10 \cite{darlow2018cinic}, CIFAR-10 \cite{krizhevsky2009cifar}, and CIFAR-100 \cite{krizhevsky2009cifar}) and three medical image datasets (OrganAMNIST \cite{yang2023medmnist}, PathMNIST \cite{yang2023medmnist}, TissueMNIST \cite{yang2023medmnist}). We utilized LeNet \cite{aggarwal2025fl} for the single-channel datasets (OrganAMNIST and TissueMNIST), whereas ResNet-18 \cite{lee2022preservation,lu2023federated} was adopted for the remaining datasets. Following prior work \cite{lu2023federated, lee2024fedsol,lee2022preservation}, clients trained 5 local epochs per round using SGD with a momentum coefficient of 0.9 and weight decay of $1\times10^{-5}$. The batch size was set to 128. After each round, the learning rate was decayed exponentially with a factor of 0.99.  To ensure sufficient convergence, we set the total training rounds to 300 for OrganAMNIST and 1000 for all remaining datasets. By default, one anchor per class was generated in each round.\\
\indent\textbf{Baseline Methods:} We compared FedADB with a broad set of representative baselines. Specifically, the benchmark suite includes the canonical FedAvg, representative methods for mitigating data heterogeneity, including FedDyn \cite{acar2021federated}, FedASAM \cite{caldarola2022improving}, FedGELA \cite{fan2023federated}, FedExp \cite{jhunjhunwala2023fedexp}, FedDecorr \cite{shi2023towards}, and FedACG \cite{kim2024communication}, as well as SOTA methods for mitigating forgetting in FL, including FedNTD \cite{lee2022preservation}, FedLMD \cite{lu2023federated}, and FedSOL \cite{lee2024fedsol}. All methods were implemented in PyTorch (v2.2.1, CUDA 12.1) and evaluated on an Intel Xeon Platinum 8375C CPU (2.90 GHz) with an NVIDIA RTX 3090 GPU.
\\
\indent\textbf{Scenario.} We evaluated all methods under two standard FL scenarios: cross-silo and cross-device\cite{chen2024free, liu2024cross, li2024incentive}. We fixed the total number of clients to $K=20$ with a participation rate of $\rho=0.2$ for the cross-silo setting, and $K=100$ with $\rho=0.05$ for the cross-device setting. We adopted the Latent Dirichlet Allocation (LDA) scheme to simulate non-IID data distributions\cite{lee2022preservation, lu2023federated}, where $\alpha$ controls the degree of cross-client data heterogeneity.
\subsection{Performance under Data Heterogeneity}
We evaluated FedADB against all baselines under both cross-silo and cross-device settings, with the results summarized in Tab.\ref{tab:main_results}. We report the mean $\pm$ standard deviation of the top-5 round accuracies across three independent trials with different random seeds. While baselines manifest sensitivity to specific configurations and occasionally fall below FedAvg, FedADB consistently outperforms SOTA methods in nearly all settings and achieves the best results in most cases. Across all datasets and non-IID settings, FedADB improved the average accuracy over the second-best method by 6.78\% and 4.71\% in cross-device and cross-silo scenarios, respectively. Compared to canonical FedAvg, the performance gains reached 13.94\% and 7.13\%, respectively. Notably, under the most severe Non-IID condition ($\alpha = 0.05$) where catastrophic forgetting is most pronounced, FedADB surpassed the second-best method by 7.52\% and 13.89\%. These results demonstrate that FedADB remains effective under extreme data heterogeneity, stabilizing global model performance, and effectively mitigating catastrophic forgetting.

\subsection{Convergence Speed Analysis}
\begin{figure*}[htbp]
    \centering
    \includegraphics[width=0.9\linewidth]{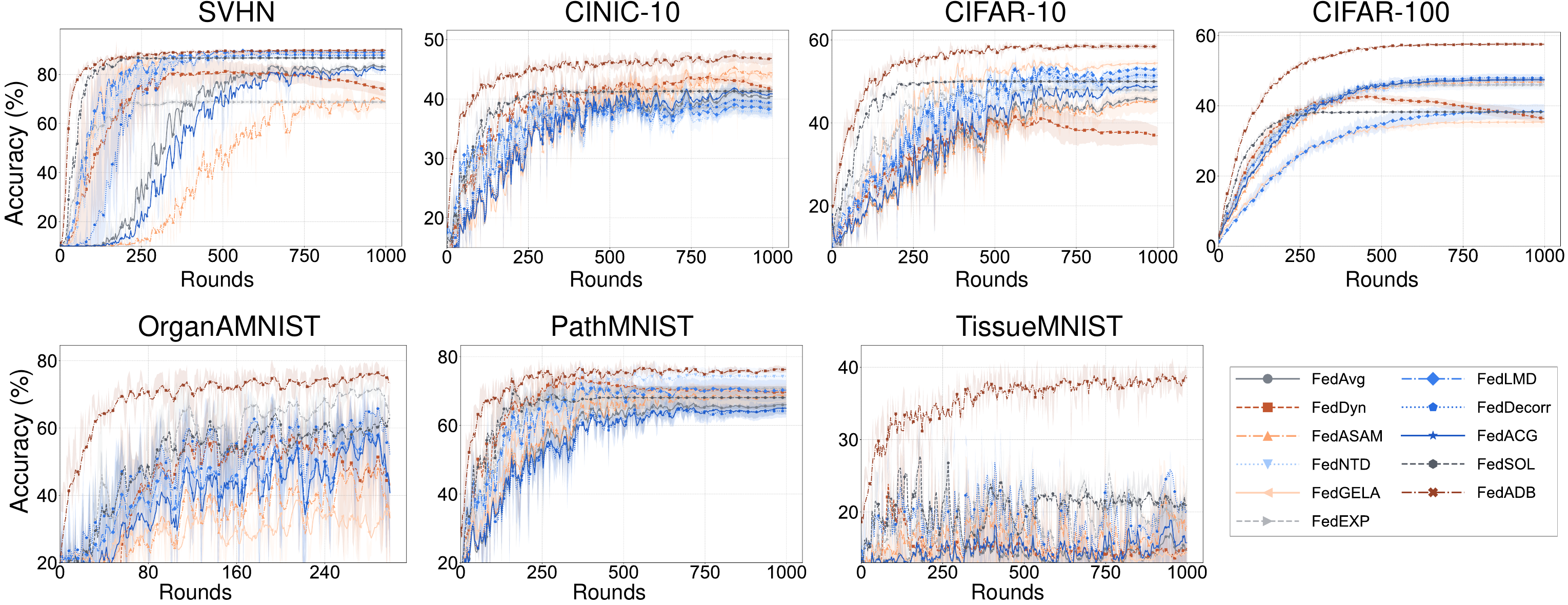} 
    \caption{Test accuracy trajectories across seven datasets ($\alpha=0.1$). Curves are smoothed using an EWMA with factor of 0.2, and the shaded areas represent the 95\% confidence intervals based on three random seeds.}
    \Description{See caption.}
    \label{fig:exp1}
\end{figure*}

Following previous works\cite{gao2024fedsts, wu2024fedekt, zhang2024multi, luo2023gradma}, we measured convergence speed by the number of communication rounds required to reach a target accuracy $T_{\text{Acc}}$. To suppress spurious spikes, we smoothed each accuracy trajectory using an Exponentially Weighted Moving Average (EWMA)\cite{hunter1986exponentially} with a factor of 0.2. We set the maximum accuracy as a reference and recorded the minimum rounds needed to reach 45\%, 65\%, and 85\% of this threshold. The cross-device results are summarized in Tab. \ref{tab:Convergence}. FedADB exhibited markedly faster convergence than all baselines: on SVHN, it reached 80\% accuracy in 50 rounds, whereas FedAvg required 478 rounds ($9.56\times$ longer). Under more challenging heterogeneous conditions (CIFAR-10 with $\alpha=0.05$), FedADB reached 35\% accuracy in 63 rounds, while other methods required 323 rounds on average ($5.13\times$ longer). Across all datasets, FedADB achieved the target accuracy with 74.4\% and 86.5\% fewer rounds than the second-best ( $3.9\times$ longer) and FedAvg ( $7.4\times$ longer), respectively. Notably, methods failing to reach the $T_{\text{Acc}}$ were excluded from the average, suggesting that FedADB's advantages are even more pronounced. As shown in Fig.~\ref{fig:exp1}, FedADB exhibits more stable optimization and faster convergence, benefiting from effective global knowledge retention and local adaptation.
% Convergence trajectories and summary statistics for cross-silo scenarios are provided in Appendix B.

\begin{figure}[t]
    \centering
    \includegraphics[width=\linewidth]{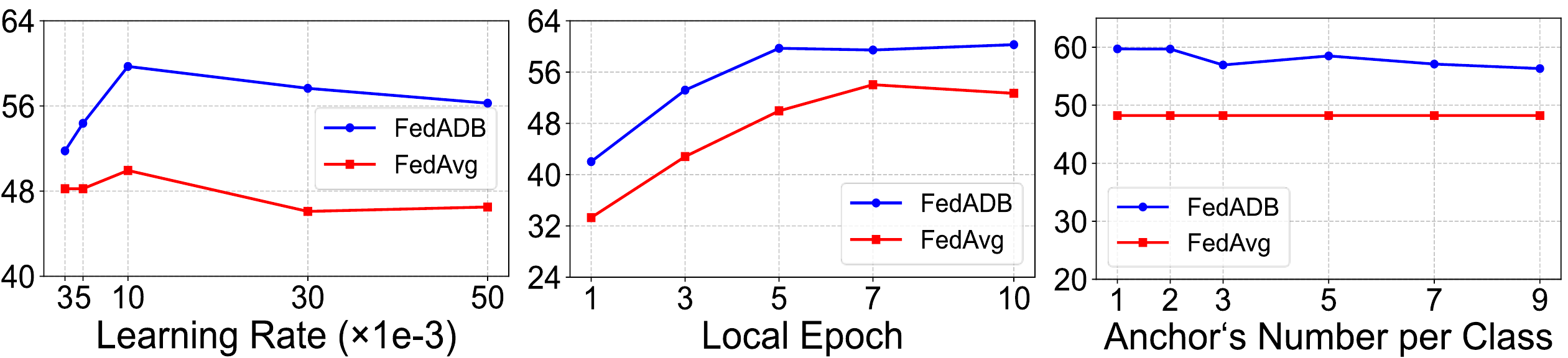} 
    \caption{ Performance of FedADB and FedAvg on CIFAR-10 ($\alpha=0.1$) across varying FL hyperparameter settings.}
        \Description{See caption.}
    \label{fig:Hyperparameters}
\end{figure}

\subsection{Hyperparameter Sensitivity }
We evaluated the sensitivity of FedADB to key FL hyperparameters, including the number of anchors per class, local epochs, and learning rate. Figure~\ref{fig:Hyperparameters} shows that FedADB consistently outperforms FedAvg across all evaluated settings. Moreover, FedADB is insensitive to the number of anchors per class, exhibiting only minor variations in accuracy. Although increasing local epochs or changing learning rate could improve the performance, FedADB maintains a distinct advantage throughout the evaluated range. Overall, these results confirm the robustness and tunability of FedADB.

\begin{figure}[h]
    \centering
\includegraphics[width=1.0\linewidth]{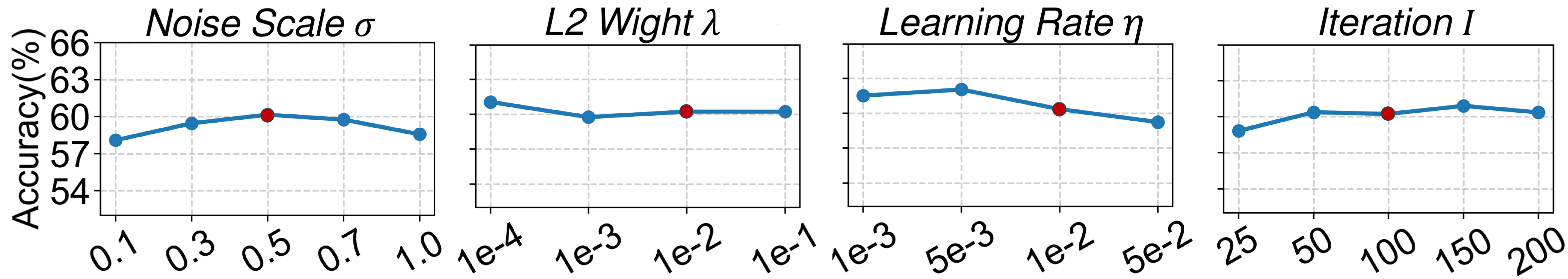}  
    \caption{ Performance of FedADB under different hyperparameters settings of CAG}
        \Description{See caption.}
    \label{fig:HyperCAG}
\end{figure}

We further evaluated the sensitivity of FedADB to the key CAG hyperparameters in Fig.~\ref{fig:HyperCAG}. The default values in our experiments, indicated by red dots, consistently lie within high-performing regions. Overall, FedADB exhibits minimal performance fluctuations across diverse CAG configurations, further confirming its stability.

\subsection{Ablation Study }

To assess the contribution of each component, we conducted an ablation study on CIFAR-10 under three non-IID settings $\alpha\in\{0.05,0.1,0.2\}$, as summarized in Tab.\ref{tab:Ablation}. First, combining AGB with LCB (w/o CAG) yielded absolute gains of +8.20\%, +2.64\%, and +1.41\%, respectively. This demonstrates that the dual-branch architecture effectively balances local learning and global knowledge preservation, whose critical role is highlighted by the more pronounced gains under severe heterogeneity ($\alpha=0.05$). Next, combining AGB with CAG (w/o LCB) produced consistent improvements of +6.53\%, +2.54\%, and +2.68\%. These gains suggest that class anchors supply explicit supervision for missing or underrepresented classes, which is increasingly valuable under severe heterogeneity. Furthermore, integrating all components yielded the largest performance boosts of +16.19\%, +11.53\%, and +8.52\%, highlighting the importance of jointly leveraging global knowledge retention, local adaptation, and anchor-guided supervision. Crucially, the overall gains surpass the sum of individual contributions, further validating the structural complementarity of FedADB.

\begin{table}[t]
    \centering
    \small
    \caption{Ablation study for different components of FedADB.}
    \label{tab:Ablation}
    \setlength{\tabcolsep}{4.5pt}
    \renewcommand{\arraystretch}{0.9}
    \begin{adjustbox}{max width=\columnwidth}
    \begin{tabular}{ccc|ccc}
        \toprule
        \multicolumn{2}{c}{\textbf{DBCT}}
        & \multirow{2}{*}{\textbf{CAG}}
        & \multicolumn{3}{c}{\textbf{Test accuracy (\%)}} \\
        \cmidrule(lr){1-2}
        \cmidrule(lr){4-6}

        \textbf{AGB}
        & \textbf{LCB}
        &
        & \boldmath$\boldsymbol{\alpha=0.05}$
        & \boldmath$\boldsymbol{\alpha=0.1}$
        & \boldmath$\boldsymbol{\alpha=0.2}$ \\
        \midrule

        $\times$
        & $\times$
        & $\times$
        & 33.87
        & 47.68
        & 56.03 \\

        $\checkmark$
        & $\checkmark$
        & $\times$
        & 42.07\scriptsize{(+8.20)}
        & 50.32\scriptsize{(+2.64)}
        & 57.44\scriptsize{(+1.41)} \\

        $\checkmark$
        & $\times$
        & $\checkmark$
        & 40.40\scriptsize{(+6.53)}
        & 50.22\scriptsize{(+2.54)}
        & 58.71\scriptsize{(+2.68)} \\

        $\checkmark$
        & $\checkmark$
        & $\checkmark$
        & \textbf{50.06}\scriptsize{(+16.19)}
        & \textbf{59.21}\scriptsize{(+11.53)}
        & \textbf{64.55}\scriptsize{(+8.52)} \\

        \bottomrule
    \end{tabular}
    \end{adjustbox}
\end{table}
We further conducted an ablation study in Tab.~\ref{tab:beta_ablation} to verify the dynamic balancing coefficient $\beta$. Compared to fixed branch weights ranging from 0.1 to 0.9, the dynamic $\beta$ achieved superior performance on both CIFAR-10 and CIFAR-100. Although fixed weights also bring certain gains, $\beta$ provided a more optimal balance by dynamically adjusting branch contributions based on data heterogeneity and anchor quality, ultimately leading to better performance.

\begin{table}[t]
    \centering
    \caption{Ablation study of fixed branch weights and the proposed dynamic balancing coefficient $\beta$.}
    \label{tab:beta_ablation}

    \setlength{\tabcolsep}{5pt}
    \renewcommand{\arraystretch}{0.85}

    \begin{adjustbox}{max width=\columnwidth}
    \begin{tabular}{lcccccc}
        \toprule
        \textbf{Dataset}
        & \textbf{0.1}
        & \textbf{0.3}
        & \textbf{0.5}
        & \textbf{0.7}
        & \textbf{0.9}
        & \textbf{$\beta$} \\
        \midrule

        CIFAR-10
        & 56.95
        & 55.32
        & 55.09
        & 52.08
        & 52.65
        & \textbf{59.21} \\

        CIFAR-100
        & 56.03
        & 54.40
        & 56.03
        & 56.64
        & 56.61
        & \textbf{57.70} \\

        \bottomrule
    \end{tabular}
    \end{adjustbox}
\end{table}

\section{Analysis}

\begin{figure}[h]
    \centering
    \includegraphics[width=\linewidth]{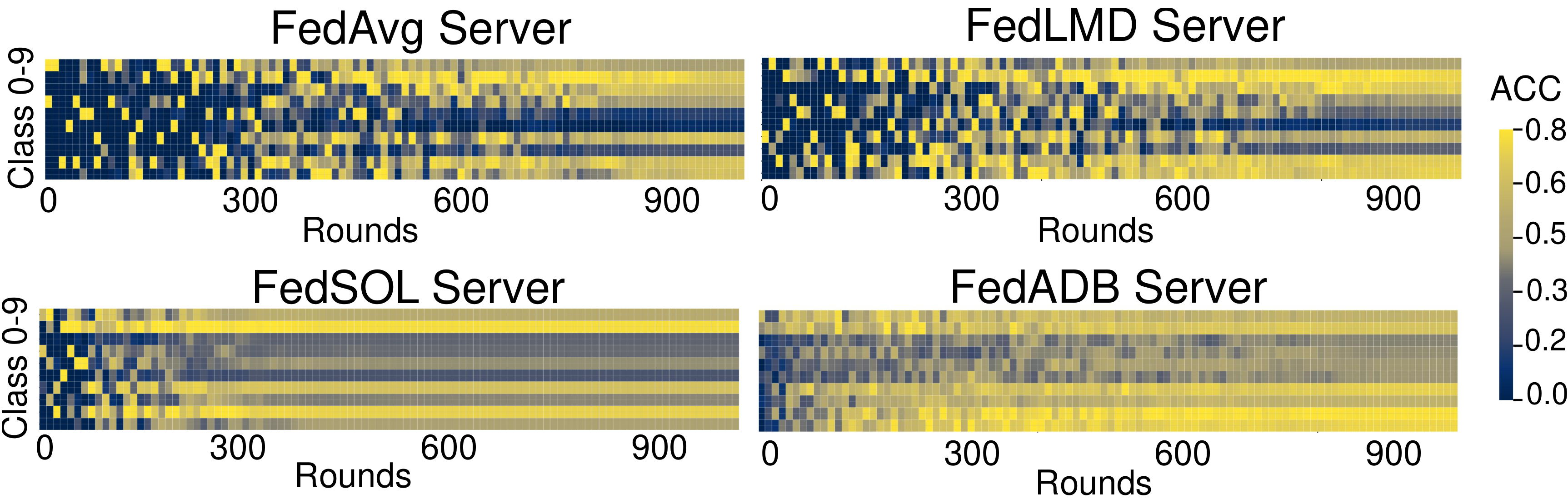}  
    \caption{Class-wise accuracy of the global model over rounds. }
        \Description{A heatmap showing the evolution of class-wise test accuracy of the global model over communication rounds. Each row represents a class, each column represents a communication round, and the color intensity indicates the corresponding test accuracy.}

    \label{fig:hotmap}
\end{figure}

\paragraph{\textbf{Knowledge Preservation}} We systematically evaluated the ability of FedADB to preserve global knowledge throughout federated training. Fig. \ref{fig:hotmap} shows the evolution of class-wise accuracy on CIFAR-10 with $\alpha=0.1$, with lighter colors indicating higher accuracy. FedAvg exhibits periodic collapse and inter-round instability on minority classes: after each round of local updates, the performance on several classes drops sharply and fluctuates dramatically. Although fluctuations persist, FedLMD and FedSOL improve knowledge retention relative to FedAvg. In contrast, FedADB maintains higher and more balanced class-wise accuracy, effectively suppressing inter-round oscillations. These results demonstrate that FedADB consistently preserves and propagates global knowledge, resulting in more stable convergence and better performance.

\begin{figure}[h]
    \centering
    \includegraphics[width=\linewidth]{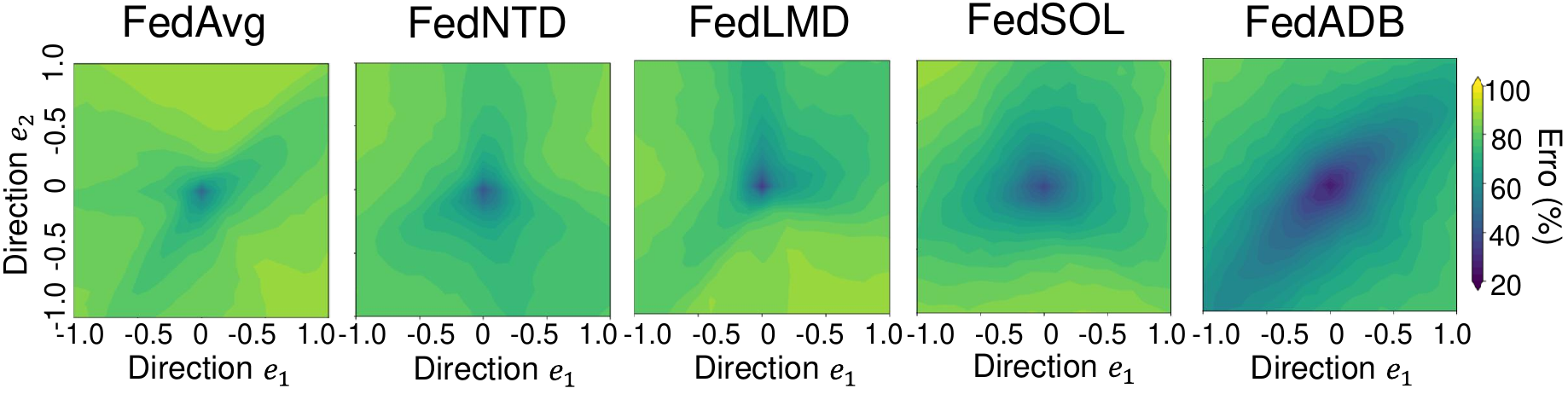}  
    \caption{Loss landscapes of converged global models projected onto the two dominant Hessian eigendirections.}
    \label{fig:landscape}
    \Description{See caption.}
\end{figure}

\paragraph{\textbf{Loss Landscape}} Prior studies indicate that models converging to flatter minima are more robust in retaining previously learned knowledge when adapting to new data distributions \cite{chen2020simple, chen2021empirical}. We analyze the global models using Hessian eigen-decomposition \cite{konevcny2016federated}, computing the top-5 eigenvalues via power iteration at each converged solution. Fig.~\ref{fig:landscape} visualizes the loss landscapes along the two principal eigendirections, while Tab.~\ref{tab:landscape} quantifies $\lambda_1$ (sharpness), $\lambda_1/\lambda_5$ (anisotropy), and training accuracy ( knowledge retention).

\begin{table}[h]
    \centering
\caption{Hessian spectrum characteristics and training accuracy of the global models on CIFAR-10 with $\alpha=0.1$.}
    \label{tab:landscape}
    \setlength{\tabcolsep}{4.5pt}
    \renewcommand{\arraystretch}{1.05}

    \begin{adjustbox}{max width=\columnwidth}
    \begin{tabular}{lccccc}
        \toprule
        \textbf{Metric}
        & \textbf{FedAvg}
        & \textbf{FedNTD}
        & \textbf{FedLMD}
        & \textbf{FedSOL}
        & \textbf{FedADB} \\
        \midrule

        $\lambda_1$
        & 311
        & 169
        & 298
        & 126
        & \textbf{99} \\

         $\lambda_1/\lambda_5$
        & 2.41
        & 1.95
        & 1.47
        & 1.46
        & \textbf{1.23} \\

        Accuracy 
        & 52.27
        & 57.48
        & 66.23
        & 62.66
        & \textbf{74.78} \\

        \bottomrule
    \end{tabular}
    \end{adjustbox}
\end{table}

Notably, FedADB converges to the flattest and most isotropic region (lowest $\lambda_1$ and $\lambda_1/\lambda_5$), yielding the highest accuracy (74.78\%). Conversely, FedAvg suffers from severe directional non-uniformity (highest $\lambda_1=311$ and $\lambda_1/\lambda_5=2.41$), and the lowest accuracy (52.27\%). This quantitative analysis validates the intrinsic relationship between landscape flatness and knowledge retention, further demonstrating FedADB's substantial advantage in constructing smooth optimization landscapes to preserve global knowledge.

\begin{figure}[h]
    \centering
    \includegraphics[width=\linewidth]{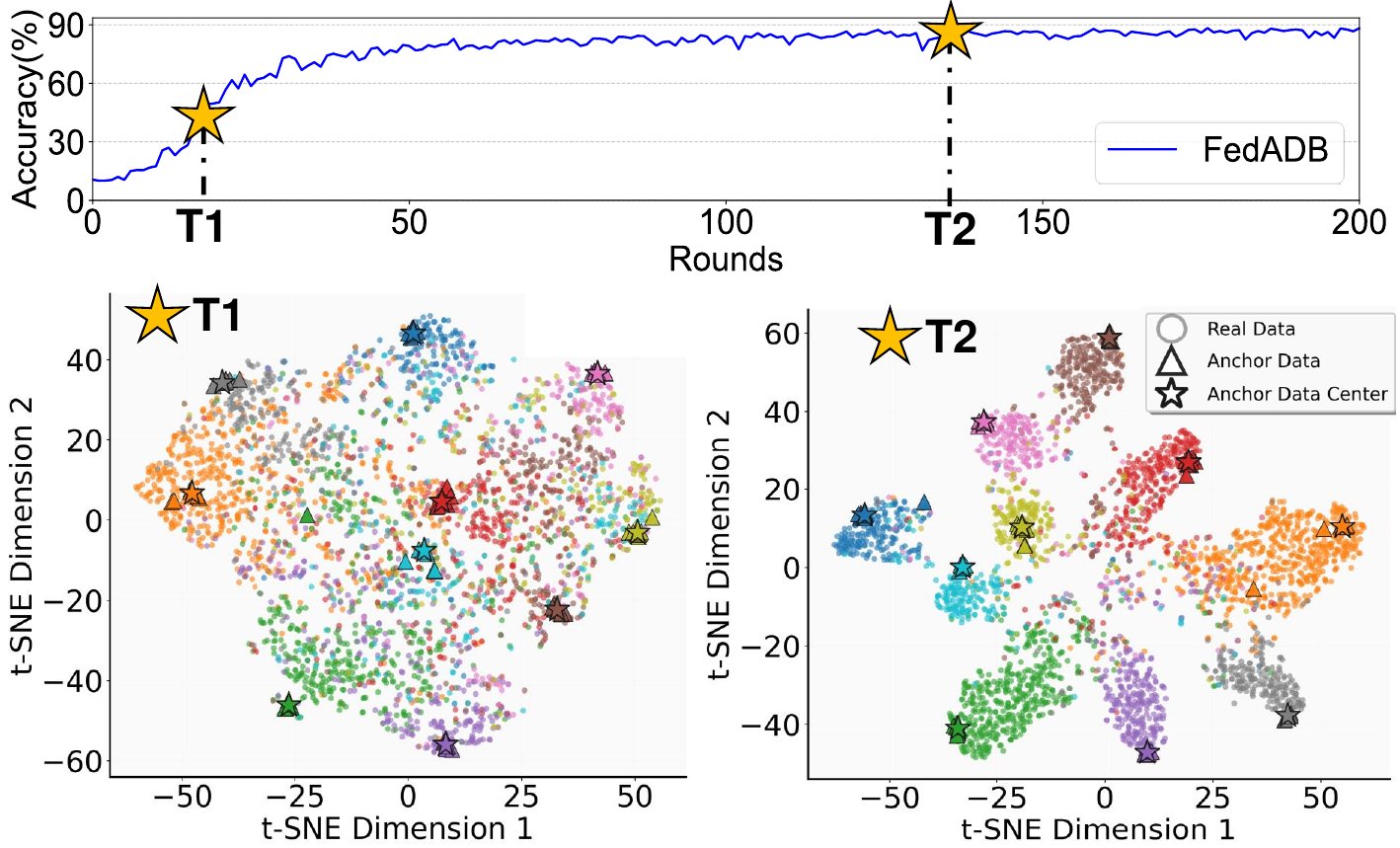}  
    \caption{t-SNE visualization of class anchors and real-data representations at different training stages on SVHN.}
        \Description{Two t-SNE plots compare class anchors with real-data representations at the early and late training stages. At the later stage, anchors move closer to their corresponding class clusters, while different classes become more clearly separated.}

    \label{fig:tsne}
\end{figure}

\paragraph{\textbf{Class Anchor Prototype Alignment}} We investigate whether class anchors align with the class-semantic prototypes represented by the evolving global model. Fig.~\ref{fig:tsne} presents the t-SNE embeddings of class anchors and real-data on SVHN with $\alpha=0.1$. From early (T1) to late (T2) stages, anchors progressively converge toward their corresponding real-data cluster centers. Simultaneously, inter-class margins expand, yielding more pronounced cluster separation. By T2, anchor embeddings substantially overlap with the real-data distribution. These observations suggest that class anchors serve as consistent global references across clients, preserving inter-class separability and decision boundaries during local training.

\begin{figure}[h]
    \centering
    \includegraphics[width=1\linewidth]{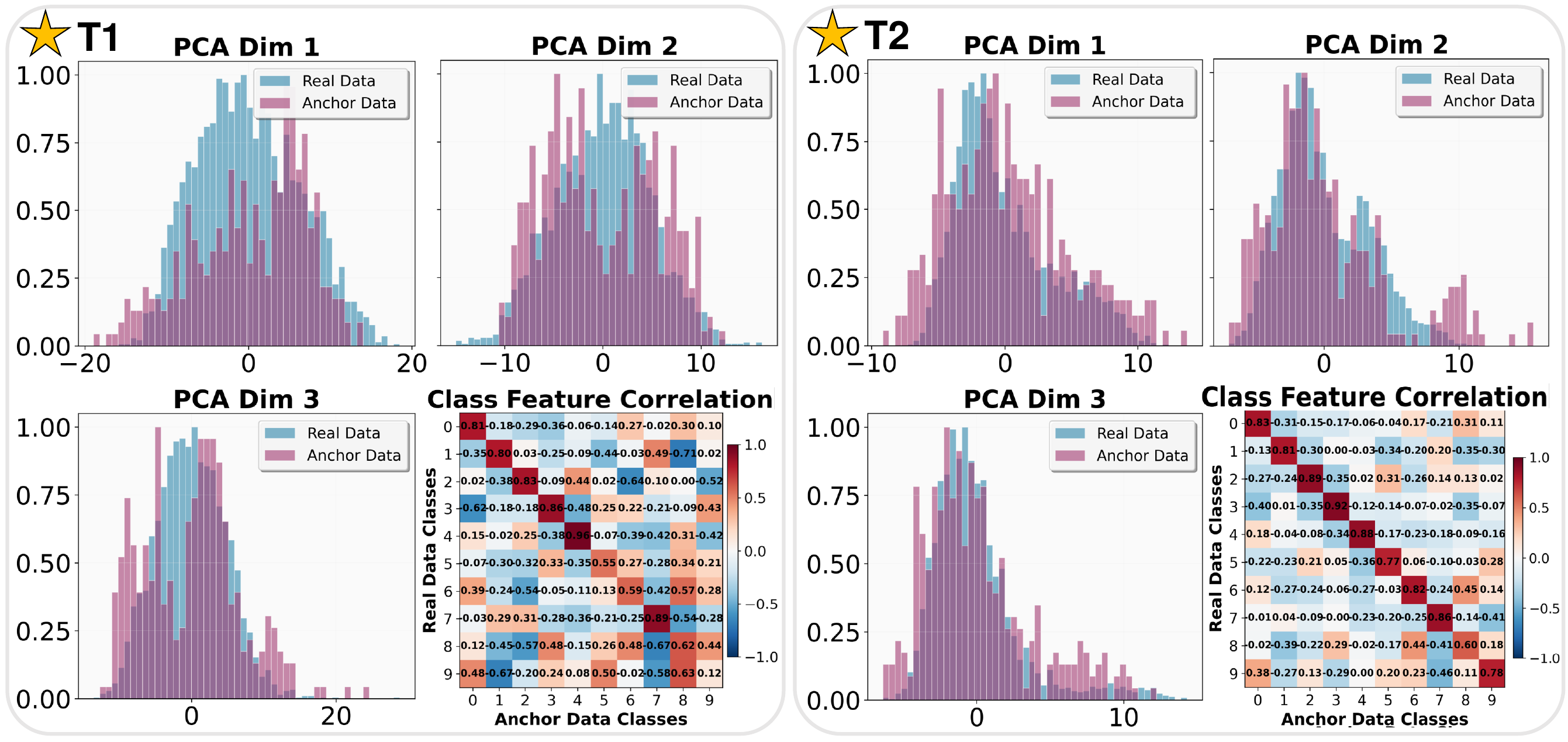} 

    \caption{PCA-based feature distributions and correlation analysis between class anchors and real-data class prototypes at  different training stages on SVHN.}
        \Description{PCA feature distributions and correlation matrices compare class anchors with real-data class prototypes at two training stages. From the early to the late stage, within-class correlations increase, off-diagonal correlations decrease, and class-wise feature alignment becomes clearer.}

    \label{fig:pca}

\end{figure}

We further quantify this alignment statistically. Specifically, we project features onto the first three principal components via PCA and compute the Pearson correlation coefficients between class anchors and real-data class centers. As shown in Fig.~\ref{fig:pca}, anchor representations increasingly align with real-data semantics: diagonal correlations are progressively strengthened while off-diagonal correlations are suppressed. This trend demonstrates improved class-wise correspondence and reduced inter-class confusion, confirming that the synthesized anchors effectively capture the semantic prototypes, corroborating the t-SNE visualizations.
% Additional visualizations of class-anchor prototypes at four stages are provided in Appendix C.

\paragraph{\textbf{Cross-Domain Dataset}} To further assess the robustness of FedADB against severe feature shifts and label skew, we conduct experiments on the PACS cross-domain benchmark \cite{li2017deeper} under a cross-silo setting with label-space mismatch. As reported in Tab.~\ref{tab:cross_domain}, FedADB yields a substantial performance margin of 7.59\% over the strongest baseline. This  demonstrates FedADB's superior capability in generalizing across diverse visual domains.

\begin{table}[h]
\centering
\normalsize
\setlength{\tabcolsep}{2.3pt}
    \renewcommand{\arraystretch}{1.05}
\setlength{\aboverulesep}{1.2pt}
\setlength{\belowrulesep}{1.2pt}

\caption{Accuracy comparison on PACS with InceptionV1}
\begin{tabular}{l|ccccc}
\toprule
Method & FedAvg & FedNTD & FedLMD & FedSOL & FedADB \\
\midrule
Accuracy & 40.35{\scriptsize (±0.13)}   & 44.29{\scriptsize (±1.54)} & 48.71{\scriptsize (±1.03)} & 43.08{\scriptsize (±0.44)} & \textbf{56.30}{\scriptsize (±0.58)} \\
\bottomrule
\end{tabular}
\label{tab:cross_domain}
\end{table}

\begin{figure}[h]
    \centering
    \includegraphics[width=1\linewidth]{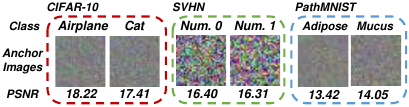}  
    \caption{Visualization and PSNR of the class anchors. }
        \Description{Examples of class anchors generated for CIFAR-10, SVHN, and PathMNIST. The anchors contain no clearly recognizable semantic details and visually resemble high-confidence noise patterns.}

    \label{fig:pse}
\end{figure}

\paragraph{\textbf{Privacy Analysis of Class Anchors}} We visualize the class anchors generated on CIFAR-10, SVHN, and PathMNIST in Fig.~\ref{fig:pse}. CAG only makes explicit the public prior knowledge in the global model. Therefore, transmitting class anchors does not introduce additional privacy leakage beyond sharing the global model itself. This privacy-preserving property is further supported by the low PSNR values and the visually unrecognizable details shown in Fig.~\ref{fig:pse}. Notably, the effectiveness of FedADB does not depend on photorealistic anchor generation. Although the anchors appear as high-confidence noise in the pixel space, they remain aligned with the cluster centers of real data in the feature space while preserving inter-class separability, as illustrated by the t-SNE visualization in Fig.~\ref{fig:tsne}. Consequently, these anchors can provide stable supervisory signals and gradient constraints for locally missing classes.

\section{Conclusion}
Federated learning is an important technical paradigm for joint analysis of multimodal data across multiple data sources. Existing FL methods for mitigating catastrophic forgetting suffer from conflicting objectives between global knowledge preservation and local adaptation, while lacking effective supervision for missing classes under non-IID settings. To address these challenges, we propose FedADB, which incorporates two key innovations. First, we construct class anchors directly from the global model through differentiable optimization, thereby eliminating the dependence on auxiliary priors and avoiding additional privacy risks. Notably, only one anchor per class is required, achieving lightweight yet effective global knowledge preservation. Second, we introduce a dual-branch collaborative training mechanism that simultaneously preserves global knowledge and learns locally discriminative features, alleviating the global-local conflict inherent in existing methods. Extensive experiments on both medical image and natural image datasets validate the effectiveness and advantages of FedADB.

\section*{Acknowledgements}

This work is supported by the National Natural Science Foundation of China (Grant No. 62472047, No. 62572068, No. 62502042, No. 62572073), and the 111 Project (Grant No. B21049).

\bibliographystyle{ACM-Reference-Format}
\balance
\bibliography{references}
\end{document}